\documentclass[12pt, draftclsnofoot, onecolumn]{IEEEtran}
\ifCLASSINFOpdf
\else
\fi
\usepackage{multirow}
\usepackage{amsbsy,amsfonts,amsfonts,amssymb,amsmath,epsfig}
\usepackage{graphicx}
\usepackage{graphics}
\input psfig
\usepackage[dvipsnames]{xcolor}
\usepackage{epstopdf}
\usepackage{array}
\usepackage{cite}
\usepackage{flushend}

\usepackage{algorithmic}
\usepackage{algorithm}

\usepackage[cmintegrals]{newtxmath}
\ifCLASSOPTIONcompsoc
\else
\fi
\begin{document}
		%
		\title{Deep-Learning Based Reconfigurable Intelligent Surfaces for Intervehicular Communication}
		%
		%
		%
		
		\author{Bulent Sagir, Erdogan Aydin and Haci Ilhan, \IEEEmembership{Senior Member,~IEEE}
			\thanks{B. Sagir is with Turk Telekomunikasyon A. S, 34660, Ist., Turkey (e-mail: bulent.sagir@turktelekom.com.tr).}
			\thanks{E. Aydin is with the Dept. of Electrical and Electronics
				Engineering, Istanbul Medeniyet Univ., Ist. 34857, Turkey (e-mail:
				erdogan.aydin@medeniyet.edu.tr) (Corresponding author: Erdogan Aydin.)}
			\thanks{H. Ilhan is with Y{\i}ld{\i}z Tech. Univ., Dept. of Electronics and Communications Engineering, 34220, Davutpasa, Ist., Turkey (e-mail: ilhanh@yildiz.edu.tr).}
		}

	\maketitle
	
	\begin{abstract}
		This letter proposes a novel deep neural network (DNN) assisted cooperative reconfigurable intelligent surface (RIS) scheme and a DNN-based symbol detection model for intervehicular communication over cascaded Nakagami-$m$ fading channels. In the considered realistic channel model, the channel links between moving nodes are modeled as cascaded Nakagami-$m$ channels, and the links involving any stationary node are modeled as Nakagami-$m$\ fading channels, where all nodes between source and destination are realized with RIS-based relays. The performances of the proposed models are evaluated and compared with the conventional methods in terms of bit error rates (BER). It is exhibited that the DNN-based systems show near-identical performance with low system complexity.

	\end{abstract}
	

	%
	\IEEEpeerreviewmaketitle
	
	
	\section{Introduction}
	%
	%
	%
	%
	\IEEEPARstart{H}{igh} penetration loss and attenuation effects resulting from walls and other obstructions in modern urban environments are some of the most compelling challenges wireless network designers should face. For achieving the required data throughput capacities in the face of these challenges, next generation wireless network architectures mostly take advantage of the latest improvements in various diversity schemes, such as antenna diversity, cooperative relaying and reconfigurable intelligent surfaces (RIS). Among these, RIS is an emerging hardware technology, comprised of multiple passive electronic devices forming a reflecting surface, altering the phase angles of the incoming electromagnetic waves and reflecting with maximum strength. Initially conceived as an active frequency selective surface for indoor environments in \cite{Hu1},\cite{Wu1}, this concept has also been adapted for outdoor scenarios and deployed on building exteriors.
	
	The potential of RIS deployment in cooperative communications is an important research topic, involving the usage of single or multiple RIS-based relays ($R$) between source ($S$) and destination ($D$) nodes. Additionally, optimizing the RIS using deep learning (DL) techniques is a novel approach already studied in various works which focus on symbol estimation \cite{Khan2020DeepLearningaidedDF}, indoor signal focusing \cite{huang2019indoor}, or securing wireless communications \cite{Yang2021}. Using DL optimized RISs is also worth considering in a cooperative configuration for intervehicular communication, which is the main topic of our study.
	
	For multi-hop relaying or cooperative systems, most of the related literature focuses on the classical fading channels, such as Rayleigh, Rician, and Nakagami, to approximate the channel model, where $S$ is stationary. However, cascaded fading distributions are already demonstrated for intervehicular communications to provide a more accurate channel model, where the $S$ and $D$ nodes are in motion \cite{Ilhan1}. The product of $K$-independent Rayleigh or Nakagami-$m$ fading channels is referred to as cascaded Rayleigh or cascaded Nakagami-$m$ fading channel. For the presented work, cascaded Nakagami-$m$ fading channels will be investigated in a cooperative system realized with deep neural network (DNN) assisted RISs. 
	
	
	\subsection{Contributions}
	In our work, DNNs have been used at the relays for optimizing the phase shifts, and at $D$ for estimating the transmitted symbols.
	We can summarize the contributions of our paper as follows.
	\begin{enumerate}
		\item In the proposed model, RISs are deployed as relays in a cooperative configuration, where signals from the multiple relays are combined at $D$, or the best performing relay can be selected with relay selection (RS) schemes.
		\item Each RIS-based $R$ is optimized with a DNN implemented in the RIS hardware.
		\item At $D$, another DNN is used for symbol estimation.
		\item Various scenarios are applied for line-of-sight (LOS) and non-line-of-sight (nLOS) links in the model with a combination of Nakagami-$m$ and cascaded Nakagami-$m$ fading channels, with path loss effects also taken into account.
		\item Performances of all scenarios in the model are analyzed and compared in terms of bit error rate (BER).
	\end{enumerate}


	
	\section{System Model}
	
	This section presents the generic model of intervehicular communication through a combination of cascaded Nakagami-$m$ and Nakagami-$m$ fading channels with DNNs deployed at $R$ and $D$.
	
\begin{figure*}[t!]
		\centering
		\includegraphics[width=0.95\textwidth]{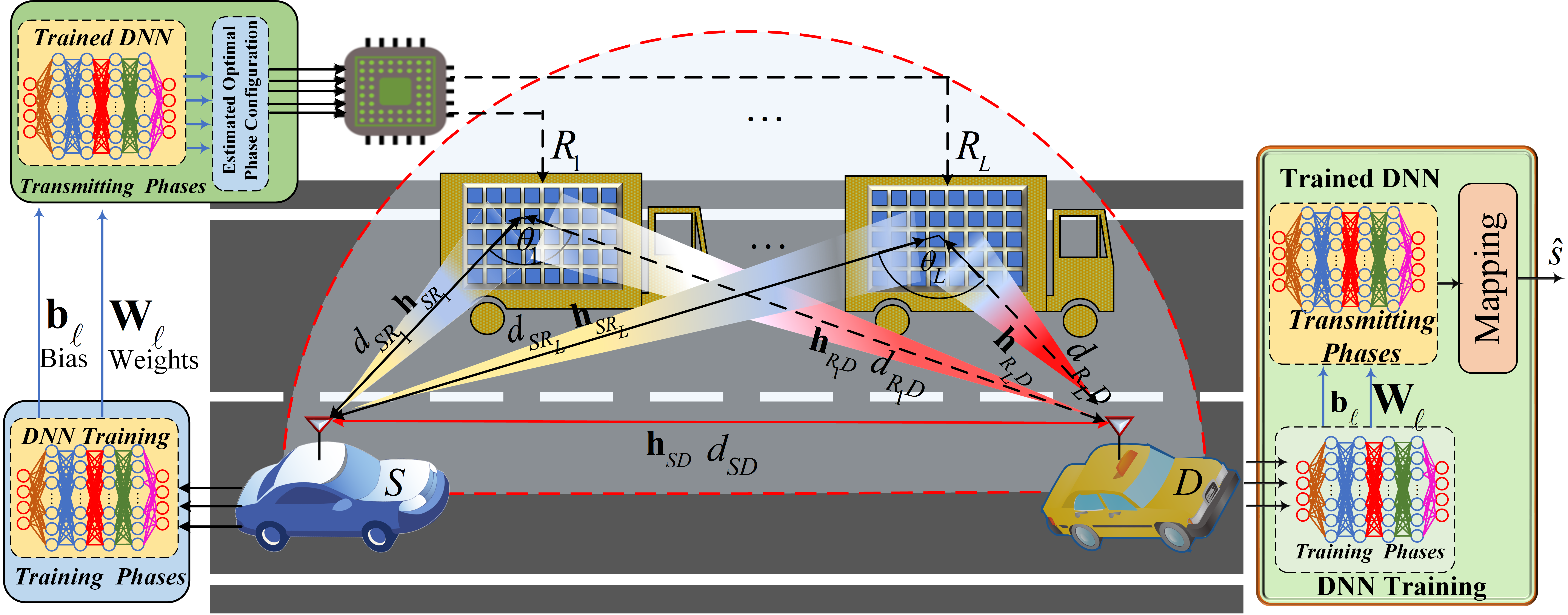}
		\caption{System model.}
		\label{org_1}
		\vspace{-0.3cm}
	\end{figure*}

	Proposed system model is shown in Fig. \ref{org_1}, where $S$, $D$, and $R$ operate in half duplex mode with a single pair of transmit and receive antennas. The distances $d_{SR_\ell}$, $d_{R_{\ell}D}$ and $d_{SD}$ represent the distances of $S$ to $\ell^{th}$ $R$ ($S\to{R_\ell}$), $\ell^{th}$ $R$ to $D$ (${R_\ell}\to{D}$) and $S$ to $D$ ($S\to{D}$), respectively. Here, $\ell=1,2,...,L$ where $L$ represents the number of relays in the system. In this configuration, $\theta_\ell$ is the angle between $S\to{R_\ell}$ and ${R_\ell}\to{D}$ links, assuming $\pi/2<\theta_\ell<\pi$ for all conditions, ensuring all relays stay within the half-circle area. 
	
	In the proposed model, we investigate vehicle-to-vehicle communications where $S$ and $D$ vehicles communicate over a single LOS link assisted with nLOS links thanks to other vehicles in their vicinity or stationary roadside access points (AP) acting as relays in a cooperative scheme. For this configuration, we evaluate two main scenarios: In the first scenario, it is assumed that all the channels involved in the transmission are subject to cascaded Nakagami-$m$ fading, where the relaying is carried out by neighboring vehicles. In the second scenario, only the channel between $S$ and $D$ vehicles is modeled by cascaded Nakagami-$m$ fading, whereas the remaining channels involving the stationary APs are modeled by Nakagami-$m$ fading.
	We assume that each RIS is formed as a reflect-array incorporating $N$  reconfigurable reflecting elements and used as a relay directing the signals from $S$ to $D$ in a cooperative network. For all the links, the path loss effect is also taken into account, which is proportional to $d^{-c}$ where $d$ is the propagation distance and $c$ is the path loss exponent. In the proposed model, the distance for the $S \to D$ link is conceived as unity, and the relative gains for $S \to R_\ell$ and $R_\ell \to  D$ links can be expressed as $\mathcal{A}_{SR_\ell}=(d_{SD}/d_{SR_\ell})^c$ and $\mathcal{A}_{R_{\ell}D}=(d_{SD}/d_{R_{\ell}D})^c$ respectively, where $\mathcal{A}_{SR_{\ell}}$ and $\mathcal{A}_{R_{\ell}D}$ are correlated by law of cosines \cite{Sagir_IoT}. As $d_{SR_{\ell}}$ and $\theta_{\ell}$ are given and $d_{SD}=1$, $d_{R_{\ell}D}$ can be calculated and path loss can be applied to channel coefficients for $S \to R_{\ell}$ and $R_{\ell} \to D$ links. Thus, the channel vectors can be depicted as  $\textbf{h}_{SR_\ell}= \big[h_{SR_{\ell,1}}, h_{SR_{\ell,2}},\ldots,h_{SR_{\ell,N}}\big] \in \mathbb{C}^{1 \times N}$ and $\textbf{h}_{R_{\ell}D}= \big[h_{R_{\ell,1}D}, h_{R_{\ell,2}D},\ldots,$ $h_{R_{\ell,N}D}\big] \in \mathbb{C}^{1 \times N}$, where $\ell=1,2,\ldots,L$, and $\mathbb{C}$ represents the set of complex numbers. Here, it is possible to express the elements of the channel vectors for both channels in terms of channel amplitudes and phases,  such as $ h_{SR_{i,j}} = \alpha_{i,j}e^{-j\varphi_{i,j}}$ and  $ h_{R_{i,j}D} = \beta_{i,j}e^{-j\theta_{i,j}}$, respectively,  where $i\in\big\{ 1,2,\ldots,L \big\}$ and $j\in\big\{ 1,2,\ldots,N \big\}$. In the first scenario, for $S \to D$, $S \to R_{\ell}$ and $R_{\ell} \to D$ cascaded Nakagami-$m$ channels, the complex channel coefficients can be expressed as the product of statistically independent, but not necessarily identically distributed, two or more Nakagami random variables such that, $h_{SD}=h_{SD_1}h_{SD_2}...h_{SD_K}$, $h_{SR_{\ell}}=h_{{SR_{\ell}}_1}h_{{SR_{\ell}}_2}...h_{{SR_{\ell}}_K}$ and $h_{R_{\ell}D}=h_{{R_{\ell}D}_1}h_{{R_{\ell}D}_2}...h_{{R_{\ell}D}_K}$, respectively, where the number of Nakagami random variables ($K$) determine the degree of cascading. As all the respective channel magnitudes follow the cascaded Nakagami-$m$ distribution, 
	the probability density function (pdf) can be expressed as,
	\begin{equation}\label{eq_pdf_cascNkgm} 
		f_h(h)=\frac{2}{h\prod_{i=1}^{K} \Gamma(m_i)}G^{K,0}_{0,K}\bigg(h^2\prod_{i=1}^{K}\frac{m_i}{\Omega_i}\bigg|^-_{m_1,...,m_K}\bigg)
	\end{equation}
	where the subscripts $SD$, $SR_{\ell}$ and $R_{\ell}D$ are omitted for convenience. Here, $G^{K,0}_{0,K}(.|:)$ is the Meijer $G$-function and $\Gamma(.)$ is the Gamma function  \cite{Ilhan1}. $m_i$ is a parameter for the fading intensity given by $m_i=\Omega^2_i/E[(h^2_i-\Omega_i)^2]\geq1/2$, with $\Omega_i=E[h^2_i]$ and $E[.]$ denoting the expectation operator.
	
	In the second scenario, as the links $S \to R_{\ell}$ and $R_{\ell} \to D$ subject to Nakagami-$m$ fading $({h_{SD}}_i, {h_{SR_{\ell}}}_i, {h_{R_{\ell}D}}_i,  i\in\{ 1,2,\ldots,K \})$, the pdf for the respective channel coefficients is $f_h(h)=\frac{2m^m}{\Omega^m\Gamma(m)}h^{2m-1}\exp\big(-\frac{m}{\Omega}h^2\big)$.

	The noisy received signal at $D$ through $S\longrightarrow{D}$ cascade LOS link only, can be expressed as
	\begin{equation}\label{eq_y_los} 
		r_{D,1} = {h}_{SD}s + n_1,
	\end{equation}
	where $s$ is the $M$-QAM modulated signal, $n_1 \sim \mathcal{CN}(0,\sigma^2 )$ is the additive white Gaussian noise (AGWN) with zero mean and $\mathcal{N}_0/2$ variance per dimension at $D$.
	The noisy baseband signals received at $D$, reflected from $R_\ell$ having $N$ reflecting elements can be given as
	\begin{eqnarray}\label{eq_y_l} 
		r_{D,2_\ell} &=& \sqrt{\mathcal{A}_{SR_{\ell}}\mathcal{A}_{R_{\ell}D}}\Bigg( \sum_{n=1}^{N}h_{SR_{\ell,n}}e^{j \phi_{\ell,n}}h_{R_{\ell,n}D} \Bigg)s + n_\ell  \nonumber \\
		&=& \sqrt{\mathcal{A}_{SR_{\ell}}\mathcal{A}_{\mathcal{A}_{\ell}D}} \bigg(\textbf{h}_{SR_\ell}\boldsymbol{\Theta}_\ell\textbf{h}_{R_{\ell}D}^T\bigg)s + n_\ell.
	\end{eqnarray}
	Here, $\boldsymbol{\Theta}_\ell \triangleq \mathtt{diag}\big(\boldsymbol{\phi}_\ell\big) \in \mathbb{C}^{N \times N}$ is the diagonal phase matrix incorporating the adjusted phase angles for each RIS reflecting element, where the phase vector for $\ell^{th}$ RIS is $\boldsymbol{\phi}_\ell=\Big[\phi_{\ell,1},\phi_{\ell,2},\dots,\phi_{\ell,N}\Big]$. The phase angles are adjusted for maximizing the signal strength at $D$, ensuring \({\phi _{\ell,i}} = {\varphi _{\ell,i}} + {\theta _{\ell,i}}\) for \(i = 1, \ldots ,N\). Finally, the vectorial representation of the noisy baseband signals reflected from all relays and received at D, can be given as 
	$\textbf{r} = [r_{D,2_1},\ldots,r_{D,2_L}]^T \in \mathbb{C}^{L \times 1}$.

	All the RISs in the system are optimized with a DNN-based algorithm, adjusting the phase angles on each reflecting element for the incoming and outgoing channel coefficients. The channel state information ($CSI$) is assumed to be perfectly known at each relay in the system, by utilizing several fixed probes on various locations for pilot signal propagation.

	\subsection{Combining Techniques at $D$}
	\subsubsection{Relay Selection (RS)}
	In the RS scheme, $R$ with the lowest BER value is selected among $L$ relays. The best relay selected for transmission can be given as
	\begin{eqnarray}\label{eq_BER_rs}
		\widetilde{BER}_{\ell^*}=\arg\min_\ell\{BER_{{D,2_1}}, BER_{{D,2_2}}, \ldots,BER_{{D,2_L}}\}.
	\end{eqnarray}
	Here, $\ell=1,2,...,L$. Also, $\ell^*$ represents the link having the lowest BER value among  $L$ relays \cite{Coop_comms_Liu}.

	\subsubsection{Maximum Ratio Combining (MRC)}
	In MRC scheme, the signals from multiple relays and $S\longrightarrow  D$ link are combined at $D$. The combined signal at $D$ can be given as
	\begin{eqnarray}\label{eq_y_MRC}
		r_{MRC}=w_0 r_{D,1} + w_1 r_{D,2_1} + w_2 r_{D,2_2}+...+w_L r_{D,2_L},
	\end{eqnarray}
	where $r_{D,1}$ and $r_{D,2_\ell}$ are defined in  (\ref{eq_y_los}) and (\ref{eq_y_l}), respectively. Here, the combining coefficients $w_0$ and $w_\ell$ are the conjugates of the complex channel coefficients to produce the sum of the absolute squared magnitudes of channel responses. Thus, the combining coefficients $w_0$ and $w_\ell$ for the $S\longrightarrow  D$ link and $R_\ell$ can be expressed, respectively, as  \cite{Sagir_IoT}
	\begin{eqnarray}\label{eq_w_l}
		w_0 &=& ({h}_{SD})^*, \nonumber \\
		w_\ell &=& \Bigg(\sum_{n=1}^N h_{SR_{\ell,n}}e^{j \phi_{\ell,n}} h_{R_{\ell,n}D}\Bigg)^* \!\!\! = \!\! \ (\mathbf{h}_{S R_\ell}\boldsymbol{\Phi}_\ell\mathbf{h}_{R_\ell D}^T)^*. 
	\end{eqnarray}
	Following the MRC-based combining at $D$, ML detector estimates the signal, which can  be defined as
	\begin{eqnarray}\label{eq_x_MRC} 
		\hat{s}_{MRC} &=& \mathrm{arg}\ \underset{\upsilon \in \{1,2,\ldots,M\}}{\min}\!\Biggl\{\Bigg|r_{MRC}\!-\!\Bigg(\!w_0{h}_{SD} +  \nonumber \\
		\! \!\! \!\! \!\! \!\! \!\! \!\! \!\! \!\! \!\! \!\! \!\! \!\! \!\! \!\! \!\! \!\! \!\! \!\! \!\! \!\! \!&  &\! \!\! \!\! \!\! \!\! \!\! \! \!\! \!\! \!\! \!\! \!\! \!\! \!\! \!\! \! \!\! \!\!\! \!+w_1\!\!\!\sum_{n=1}^{N}\!\!h_{SR_{1,n}}e^{\hat{\phi}_{1,n}}h_{R_{1,n}D}\!+...\!+\!w_L \!\!\!\sum_{n=1}^{N}\!\!h_{SR_{L,n}}e^{\hat{\phi}_{L,n}}h_{R_{L,n}D}\!\!\Bigg)\!s_\upsilon \! \Bigg|^2\Biggr\},
	\end{eqnarray}
	where $M$ represents the modulation order, and $s_v$ denotes all possible symbols to be transmitted for $v\in\{1,2,...,M\}$.
	
	
	\subsection{Proposed DNN-Based Relay and Destination Models}
	
	In this section, we will provide an overview of DNN assisted phase optimization and DNN-based symbol estimation techniques for the proposed model.
	
	As stated earlier, DNNs are implemented at two critical stages in the model. At each $R_\ell$, a DNN optimizes the phase vectors using channel coefficients for $S\longrightarrow{R_\ell}$ and ${R_\ell}\longrightarrow{D}$ links, and at $D$, another DNN estimates the symbols based on the received signals, which are exposed to path-fading and noise.
	
	In the proposed scheme, the DNN process is split into two phases: the DNNs are trained with the appropriate training set during the initial training phase, and the trained DNNs are fed with the real world data to estimate the required parameters during the transmission phase, afterward. The training data set for the DNN at $R_\ell$ comprises channel coefficients $\big(\textbf{h}_{SR_\ell},\textbf{h}_{R_\ell D}\big)$ as feature vectors and respective phase adjustments  $(\boldsymbol{\phi}_\ell\big)$ as outputs, to be processed for estimating the optimal phase vector that maximizes the signals reflected through the relays. During the transmission phase, new channel coefficients for $S \to R_\ell$ and $R_\ell \to D$ links are received by the DNN, and the DNN estimates the optimal phase vectors $(\Hat{\boldsymbol{\phi}}_\ell\big)$ and transmits them to the RIS. RIS uses these phase estimations for the current signal transmission until the channel conditions change and new channel coefficients arrive. On the other hand, the training data set for the DNN at $D$ comprises the noisy and path-faded signals received at $D$ as feature vectors and the respective modulated symbols ($s$) as outputs. During the actual transmission, the trained DNN estimates the symbols ($\hat{s}$) using the noisy signals received at $D$ in real-time.
	
	Both DNN architectures contain $4$ hidden layers, one input and one output layer, where each hidden layer has $256$ fully connected neurons. In both models, rectified linear unit (ReLU) activation function has been deployed at hidden layers, which can be expressed as $\sigma_{\text{ReLU}}(p)=\text{max}(0,p)$, for a neuron with input $p$ \cite{Khan2020DeepLearningaidedDF}.
	
	Here, defining the estimation process for the DNN at $D$ as a multiclass classification problem is critical, so that the symbol set can be mapped to a class set which will be the estimations of the DNN. On the other hand, the estimation process for the DNN at $R$ is defined as a regression problem. In this regard, the output layers of the DNN at $R$ and $D$ will be configured as regression and classification layers to estimate the complex phase adjustments ($\hat{\boldsymbol{\phi}}_\ell$) and the $M$-QAM modulated symbols ($\hat{s}$), respectively.
	
	A classification layer is typically deployed after the softmax layer, which has the activation function
	$\sigma_{\text{softmax}}(\textbf{p})_i = {e^{p_i}}/{\sum^\mathcal{K}_{j=1}e^{p_j}}$.
	Here, $\textbf{p}$ and $p_i$ represent $\mathcal{K}$ vectors from the previous layer and the input vector elements, respectively, where $i=1,...,\mathcal{K}$ and $\textbf{p}=(p_1,...,p_k)\in\mathbb{R}^\mathcal{K}$ \cite{bridle-softmax}. Softmax layer simply converts $\mathcal{K}$ vectors at the input to $\mathcal{K}$ vectors at the output where the sum of the vectors is unity. Eventually, each vector transforms into an element of a probability distribution, and the highest probability vectors are selected to be classified by the following layer.

	\subsection{Generating Training Data Set and the Training Process}

	To train the proposed DNN at $R$, we need feature vectors  $h_{SR_{\ell,n}}$ and $h_{R_{\ell,n}D}$ as input and adjusted phase angles $\boldsymbol{\phi}_{\ell}$ as output. Both input and output vectors should be separated into real and imaginary parts due to the processing constraints of the DNN, where both vectors can be defined, respectively, as
	\begin{eqnarray}\label{eq_F_train}
		\mathbf{U}_{\text{train}}^i  = \Big[ \Re\big(h_{SR_{\ell,n}} \big) \ \Im\big(h_{SR_{\ell,n}}\big) \ \Re\big(h_{R_{\ell,n}D}\big) \ \Im\big(h_{R_{\ell,n}D}\big)\Big]_{1\times4}.
	\end{eqnarray}
	\begin{eqnarray}\label{eq_O_train}
		\mathbf{G}_{\text{train}}^i  = \Big[ \Re\big(\boldsymbol{\phi}_{\ell,n}\big) \ \Im\big(\boldsymbol{\phi}_{\ell,n}\big)\Big]_{1\times2}
	\end{eqnarray}
	Here, the feature vector $\mathbf{U}_\text{train}\triangleq{\mathbf{U}_\text{train}^1,\mathbf{U}_\text{train}^2,\ldots,\mathbf{U}_\text{train}^f}$ for $f$ samples, $i=1,2,...,f$ and $\mathbf{U}_{\text{train}}^i\in\mathbf{U}_{\text{train}}$, and output vector $\mathbf{G}_\text{train}\triangleq{\mathbf{G}_\text{train}^1,\mathbf{G}_\text{train}^2,\ldots,\mathbf{G}_\text{train}^f}$ for $i=1,2,...,f$ and $\mathbf{G}_{\text{train}}^i\in\mathbf{G}_{\text{train}}$. Once we have feature and output vectors, training data set can be generated as
	\begin{eqnarray}\label{eq_D_train}
		\mathcal{D}_{\text{train}}\!=\!\big\{\!\{\mathbf{U}_\text{train}^1,\!\mathbf{G}_\text{train}^1\},\{\mathbf{U}_\text{train}^2,\!\mathbf{G}_\text{train}^2\},\dots,\{\mathbf{U}_\text{train}^f,\!\mathbf{G}_\text{train}^f\}\!\big\}.
	\end{eqnarray}
	
	The loss function for the proposed DNN architecture at $R$ can be expressed as
	\begin{eqnarray}\label{eq_loss_relay}
		\mathcal{L}(\Theta) = \frac{1}{2}\sum_{i=1}^f \left \| \boldsymbol{\phi}_{\ell,n}^i- \hat{\boldsymbol{\phi}}_{\ell,n}^i \big(\Theta\big) \right \|^2,
	\end{eqnarray}
	where, $\boldsymbol{\phi}_{\ell,n}$ is the desired output value, $\hat{\boldsymbol{\phi}}_{\ell,n}$ is the estimated output and $i$ denotes the response index. Here, $\Theta$ represents the training parameter set as $\Theta \triangleq \bigl\{\boldsymbol{\theta}_1,\boldsymbol{\theta}_2,\ldots,\boldsymbol{\theta}_{f}\bigr\}$ with $\boldsymbol{\theta}_k \triangleq \bigl\{\textbf{W}_k,\textbf{b}_k\bigr\}$,  consisting of weight and bias values of the $k^{th}$ layer, respectively.
	The loss function determines the difference between the estimated and desired output values and provides the data to be processed by the optimization algorithm to find the optimum $\textbf{W}_k$ and $\textbf{b}_k$ values on each iteration. 
	
	For the proposed DNN at $D$, the estimation problem is considered to be a multiclass classfication problem, where the estimator should converge to the nearest $\hat{s}$ for the original symbol. During the process,  $\hat{s}$ will be selected from a symbol set $\mathbb{S}$, which is basically mapped onto a set of classes $\mathcal{C}$, both of which can be defined as
	\begin{eqnarray}\label{eq_S}
		\mathbb{S}=\Big\{s_{\mathbb{S}_1},s_{\mathbb{S}_2}\ldots,s_{\mathbb{S}_M}\Big\},\mathcal{C}=\Big\{\mathbb{Y}_1,\mathbb{Y}_2,\ldots,\mathbb{Y}_M\Big\}
	\end{eqnarray}
	for $M^{th}$ modulation order.
	Here, $\mathcal{C}$ represents the set of class labels mapped to each element of the symbol set $\mathbb{S}$, where $s_{\mathbb{S}_v}\in\mathbb{S}$ and $\mathbb{Y}_v\in\mathcal{C}$ for $v=1,2,\ldots,M$. Mapping relation between $\mathbb{S}$ and $\mathcal{C}$ can be represented by  $\big\{s_{\mathbb{S}_v}\longleftrightarrow\mathbb{Y}_v\big\}\longrightarrow{\hat{s}_f}$. Finally, the feature vector and the respective output vector can be defined as
	\begin{eqnarray}\label{eq_F_train_dest}
		\mathbf{U}_{\text{train}}^i = \Big[ \Re(\mathbf{r}_{\ell,n}), \Im(\mathbf{r}_{\ell,n})\Big]_{1\times2}, \ \mathbf{G}_{\text{train}}^i  = \Big[\mathbb{Y}_v^i\Big]_{1\times1},
	\end{eqnarray}
	where $ v=1,2,\ldots,M$ and $\mathbf{r}_{\ell,n}$ is separated into real and imaginary parts as in (\ref{eq_F_train}) for $i=1,2,\ldots,f$ and $\mathbb{Y}_v$ is a class category for $\mathbf{G}_{\text{train}}^i$. Hence, the training data set can be formed as in (\ref{eq_D_train}) with feature vectors and classes paired for the sample space $f$ .
	
	The loss function for the proposed DNN at $D$ can be expressed as
	\begin{eqnarray}\label{eq_loss_dest}
		\mathcal{L}(\Theta) = -\frac{1}{f}\sum_{i=1}^f\sum_{v=1}^T \mathbb{Y}_v^i\ln{\Hat{\mathbb{Y}}_v^i}\big(\Theta\big)
	\end{eqnarray}
	Here, $f$ represents the number of samples, $T$ is the number of classes, $\mathbb{Y}_v^i$ is the desired value for the $v^{th}$ class at $i^{th}$ sample, and $\Hat{\mathbb{Y}}_v^i$ is the actual output value \cite{bridle-softmax}.
	
	
	\begin{algorithm}[t!]
		\caption{Training Algorithm for DNN @ $R_\ell$ and  @ $D$}
		\begin{algorithmic}[1]
			\renewcommand{\algorithmicrequire}{\textbf{Input:}}
			\renewcommand{\algorithmicensure}{\textbf{Output:}}
			\REQUIRE $h_{SR_{\ell,n}}$ and $h_{R_{\ell,n}D}$ for DNN @ $R_\ell$, $x$ and $\mathbf{r}_{\ell,n}$ for DNN @ $D$.
			\ENSURE  Trained DNN @ $R_\ell$ and @ D - $\textbf{W}_k$, $\textbf{b}_k$
			\\ \textit{Initialisation} : $\textbf{W}_k$, $\textbf{b}_k$, and $\mathcal{L}(\Theta)$ are set to zero.
			\\ \textit{LOOP Process}
			\FOR {$i = 1$ to $s$}
			\STATE Pre-process $h_{SR_{\ell,n}}$, $h_{R_{\ell,n}D}$ and $\mathbf{r}_{\ell,n}$, generate  $\mathbf{U}_{\text{train}}^i$ for DNN @ $R_\ell$ and @ D as in (\ref{eq_F_train}) and (\ref{eq_F_train_dest}), respectively.
			\STATE Compute $\boldsymbol{\phi}_{\ell,n}$, extract $\mathbb{S}$ and convert to $\mathcal{C}$ as in (\ref{eq_S}).
			\STATE Generate $\mathbf{G}_{\text{train}}^i$ for DNN @ $R_\ell$ and @ D as in (\ref{eq_O_train}) and (\ref{eq_F_train_dest}), respectively.
			\ENDFOR
			\STATE Generate $D_{\text{train}}$ as in (\ref{eq_D_train}) for DNN @ $R_\ell$ and @ D.
			\STATE Train DNN @ $R_\ell$ and @ D until $\mathcal{L}(\Theta)$ is minimized with respect to (\ref{eq_loss_relay}) and (\ref{eq_loss_dest}).
			\RETURN Trained DNN @ $R_\ell$ and @ $D$
		\end{algorithmic} 
	\end{algorithm}

	
	\begin{table} [t!]
		\vspace{-0.35cm}
		\caption{Training Parameters.}
		\begin{center}
			\begin{tabular}{|c|c|c|}
				\hline
				\hline
				\textbf{Training Parameter} &  \textbf{DNN@Relay} & \textbf{DNN@Destination} \\
				\hline\hline
				{Modulation Order ($M$)} &\multicolumn{2}{c|}{4} \\ 
				\hline
				{Num. of Tx and Rx Antennas} &\multicolumn{2}{c|}{1} \\
				\hline
				{Num. of Reflect. Elements ($N$)} &\multicolumn{2}{c|}{8/16/32} \\
				\hline
				{Path Loss Exponent ($c$)} &\multicolumn{2}{c|}{4} \\
				\hline
				{Num. of Hidden Layers} &\multicolumn{2}{c|}{4} \\
				\hline
				{Batch Size} &\multicolumn{2}{c|}{256} \\
				\hline
				{Validation Split} &\multicolumn{2}{c|}{10\%} \\ 
				\hline
				{Num. of Samples ($s$)} & 400000 & 50000 \\ 
				\hline
				{Iteration Steps} & 600 & 400 \\ 
				\hline
				{Learning Rate ($\eta$)} & 0.003 & 0.003 \\ 
				
				\hline
				\hline
			\end{tabular}
			\label{table:2}
		\end{center}
		\vspace{-0.4cm}
	\end{table}

	\section{Numerical Results and Discussions}
	\subsection{Simulation Parameters}
	The simulation is realized for the proposed DNN models for phase optimization at $R$ and symbol estimation at $D$. In the simulation, $S$, $R_{\ell}$, and $D$ are located as shown in Fig. \ref{org_1}, and all relays are positioned over the half circle area by the relation $ \pi/2<\theta_\ell<\pi$, where $d_{SD}$, $d_{SR_{\ell}}$, and $d_{R_{\ell}D}$ distances are normalized with respect to $d_{SD}=1$. The signal to noise ratio (SNR) used in the simulations can be given as $\mathrm{SNR (dB)}=10\log_{10}(E_s/N_0)$. All Nakagami-$m$ and cascaded Nakagami-$m$ fading channels are modeled regarding path loss effects with $c=4$, indicating a dense urban environment \cite{Rappaport1992}.
	
	Simulation is realized in the MATLAB environment using the basic training parameters given in Table \ref{table:2}. As given previously, ReLU activation function is used in hidden layers for the proposed DNN models. During training, validation split determines the percentage of training data reserved for validation process, and set as $10\%$ with the validation frequency of $10$ for both DNNs. Batch size is set as 256 for both DNN models, and defines the size of the data batch to be processed on each iteration during training.
	
	\begin{figure*}[t!]
		\centering
		\includegraphics[width=0.99\textwidth]{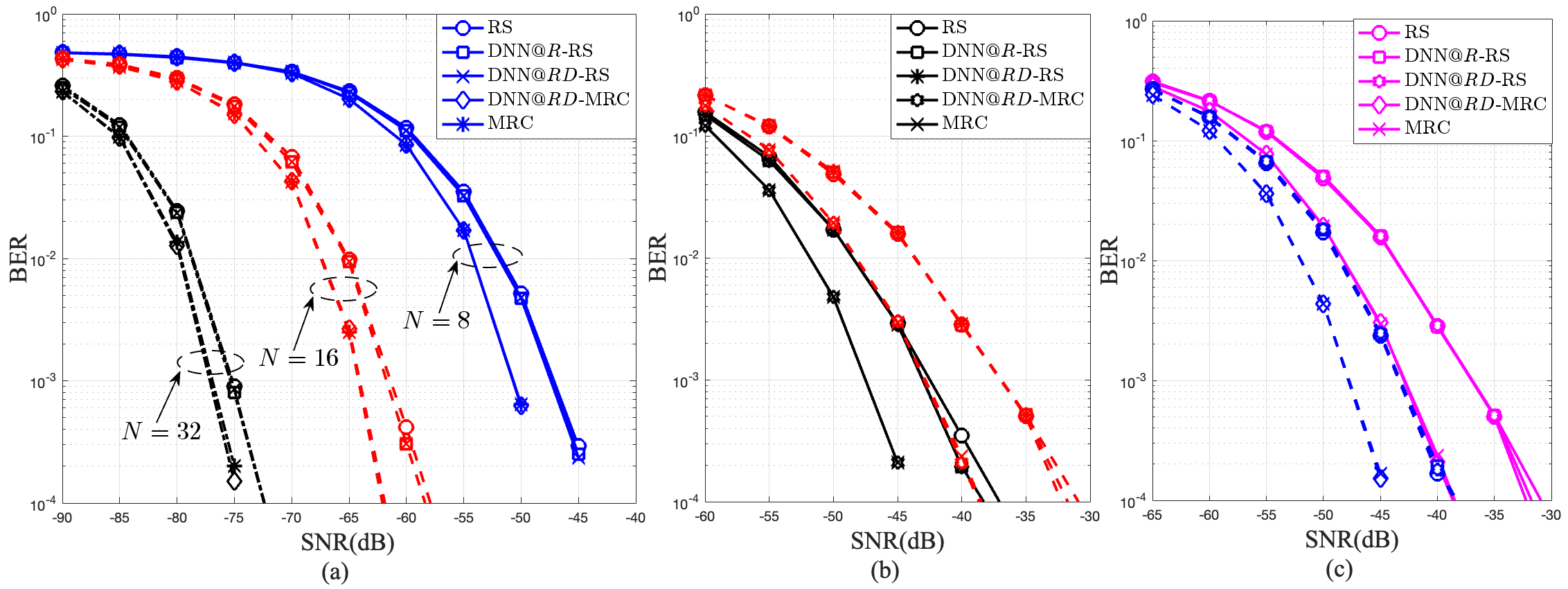}
		\caption{BER comparisons of scenario 1 schemes for (a) $N=8,16,32$, $K=2$, $m=3$, (b) $K=2$ (continuous line), $3$ (dashed line), $N=8$, $m=3$, (c) $m=2$ (continuous line), $3$ (dashed line), $K=3$, $N=8$.}
		\label{case1}
	\end{figure*}

	\subsection{BER Performance Analysis}
	
	The performances of the proposed DNN-based systems are evaluated in terms of BER using various configurations with variable $N$, $K$, and $m$ values to realize the impact of these parameters on the general performance. For all the scenarios, a two relay system is considered with $d_{SR_1}=0.5$ and $d_{SR_2}=0.7$ where RS scheme is applied at $D$ unless otherwise stated.
	
	In Fig. \ref{case1}(a), we can observe the BER performances for scenario 1 where all channels are subject to cascaded Nakagami-$m$ fading with $K=2$ and $m=3$ for 8, 16, and 32 RIS reflecting elements ($N$) with all the other parameters fixed. A huge performance gain can be achieved by increasing $N$ as expected, up to 15dB for all relay configurations. It is significant that the performance of combined DNN assisted phase and symbol estimation is nearly identical with non-DNN or relay-only DNN configurations for any RIS size.
	
	The effect of the degree of cascading is shown in Fig. \ref{case1}(b), where the performance gets worse with the increase in cascading degree ($K$), resulting with an SNR loss of nearly 6 dB per cascading at a BER level $10^{-3}$. Similarly, the effect of the Nakagami fading parameter ($m$) can be observed in Fig. \ref{case1}(c), depicting a positive correlation between BER and $m$ while the performance of the combined DNN relay and DNN destination scheme is unaffected from $K$ or $m$ variations and found to be identical with non-DNN configurations.
	
\begin{figure}[t!]
		\centering
		\includegraphics[width=0.55\textwidth]{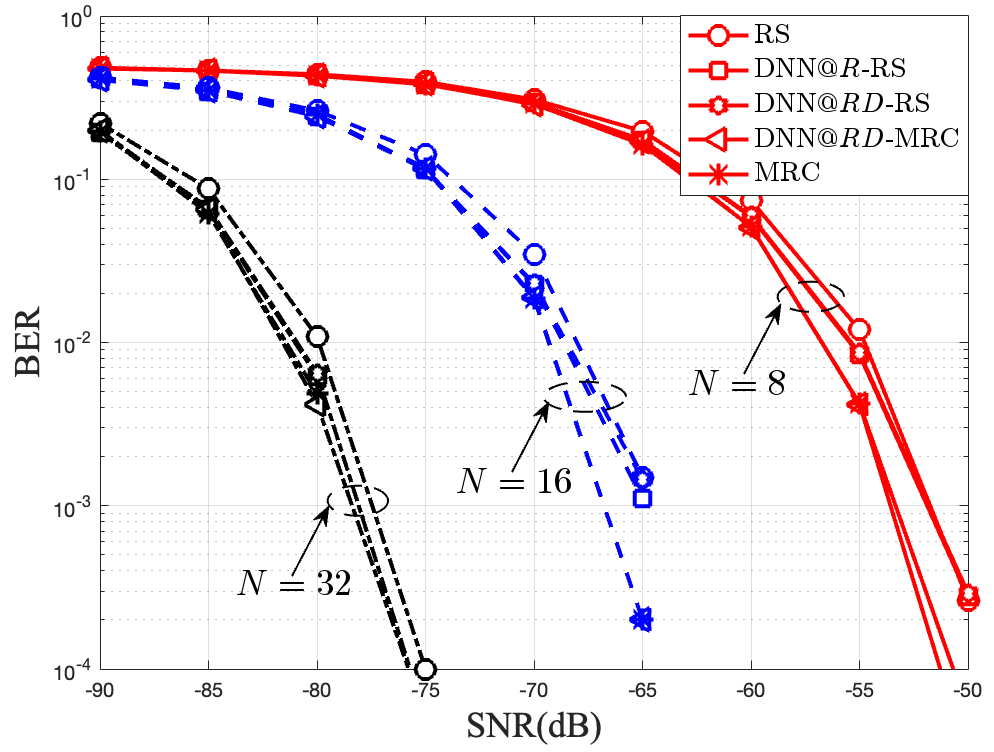}
		\caption{BER comparisons of scenario 2 schemes for $N=8,16,32$, $m=3$, $K=2$ for $S \to D$, $K=1$ for $S \to R_{\ell}$ and $R_{\ell} \to D$.}
		\label{case2}
	\end{figure}

	Finally, BER performances for the scenario 2 is shown in Fig. \ref{case2}, for 8, 16, and 32 reflecting elements. We can observe the similar SNR gains with the increase in $N$ as seen in the first scenario, but with a better overall performance as the cascading is effective only on the $S \to D$ channel. Compared to the first scenario, the performance gap between DNN-based and non-DNN-based configurations is still marginal, but the RS performances are a bit closer to MRC.

	\section{Conclusions and Future Works}
	In this paper, we have realized a DNN assisted cooperative RIS scheme with a DNN-based symbol estimator at $D$ for intervehicular communication over Nakagami-$m$ and cascaded Nakagami-$m$ fading channels. BER performances for the proposed DNN models are simulated and compared with non-DNN models, and it is found that the proposed DNN-based models for both phase and symbol estimation performed successfully in various communication scenarios, showing near-identical performance with non-DNN models.
	
	The proposed models are simulated assuming CSI is perfectly known at $R$ and $D$, exhibiting the potential of DNN deployment in intervehicular communication. However, there's an improvement area for the proposed systems against blind or imperfect CSI conditions, where more advanced DL techniques possibly be needed. Another interesting case will be the transmission of image and video over the proposed models, as the combined effect of fading, path loss, and estimation errors on the video or image signal can be a significant challenge to overcome, requiring a similar refinement in the DL approach.

	
	%

	



	\ifCLASSOPTIONcaptionsoff
	\newpage
	\fi
	
	
	%
	\vspace{-0.3cm}
	\bibliographystyle{ieeetr}
	\bibliography{Referanslar}
	
	
	
	%
	
	
	
	
	
	
	

\end{document}